# TIME DELAY ESTIMATOR FOR PREDETERMINED REPEATED SIGNAL ROBUST TO NARROWBAND INTERFERENCE

*TaeJin Park and Kyeong Ok Kang*

Electronics and Telecommunications Research Institute, Republic of Korea

## ABSTRACT

In this paper, time delay estimation techniques robust to narrowband interference (NBI) are proposed. Owing to the deluge of wireless signal interference these days, narrowband interference is a common problem for communication and positioning systems. To mitigate the effect of this narrow band interference, we propose a robust time delay estimator for a predetermined repeated synchronization signal in an NBI environment. We exploit an ensemble of average and sample covariance matrices to estimate the noise profile. In addition, to increase the detection probability, we suppress the variance of likelihood value by employing a von-Mises distribution in the time-delay estimator. Our proposed time delay estimator shows a better performance in an NBI environment compared to a typical time delay estimator.

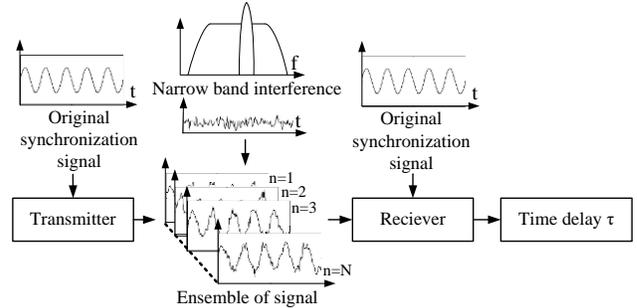

**Fig. 1.** Diagram of time delay estimation problem with predetermined signal and ensemble signal set.

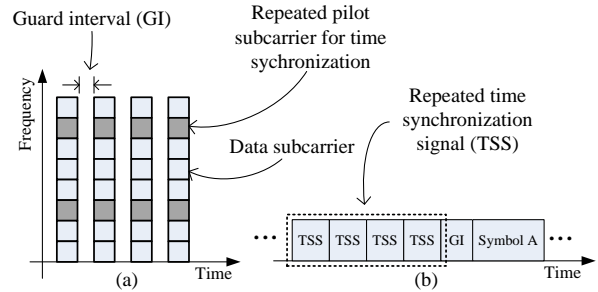

**Fig. 2.** Example of repeated time synchronization signal.

## 1. INTRODUCTION

To transfer data wirelessly such as through an orthogonal frequency division multiplex (OFDM) or code division multiple access (CDMA), the time offset estimation or time synchronization is a crucial issue because such communication systems can extract intact data only after accurately synchronizing their time offset. In addition, a global positioning system (GPS) or indoor navigation system (INS) needs an accurate time delay estimator.

A number of works related to time delay estimation have been proposed for wireless communication systems. Most of the previously developed maximum likelihood (ML) related time delay estimators (TDE) are based on the time domain. Most well-known time offset estimation techniques for OFDM are methods that jointly use an ML estimator for the cyclic prefix signal [1]. In addition, a technique employing an ML estimator for identical training symbols is widely known [2]. A maximum likelihood time delay estimator (MLTDE) using multiple identical signals and covariance was also proposed for CDMA [3] and GPS [4]. Although these methods work properly in their application, time domain based TDE is vulnerable to severe narrowband noise (NBN) or narrowband interference (NBI). Therefore, we focused on the fact that a frequency domain based ML time delay estimator can be more robust to narrowband noise than a time domain based ML estimator. A frequency domain based time delay estimator for NBI was also proposed in [5], but since its algorithm nullifies the frequency bin with a low signal-to-interference ratio (SIR), it deteriorates the detection probability since the threshold value is not able to properly filter out a frequency band with a low SIR.

In this paper, we assume that a transmitter sends multiple sets of identical known synchronization sequences for the local receiver to estimate the noise profile. These sequences can be sent by either a repeated pilot subcarrier, as described in Fig. 2 (a), or a repeated training symbol, as shown in Fig. 2 (b). Since we can estimate a noise covariance matrix from multiple sets of synchronization signals, we can effectively suppress the effect of narrowband interference by employing a frequency domain based maximum likelihood time delay estimator (FDMLTDE). Furthermore, we employ a von-Mises distribution to suppress the variance of the likelihood value. Our proposed estimator therefore shows a higher detection probability compared to a conventional correlation based method, the method proposed in [5], and other typical algorithms.

## 2. TIME DELAY ESTIMATION PROBLEM

Since we assume that an identical time synchronization signal is repeatedly transmitted from the transmitter, we have multiple signal sets with multiple subcarriers or frequencies. If we set the signal to be repeated M-times with N subcarriers (or frequencies), we can describe the k-th frequency of the M-times received repeated signal set as

$$\mathbf{Y} = \begin{bmatrix} \mathbf{y}_1[0] & \mathbf{y}_2[0] & \cdots & \mathbf{y}_M[0] \\ \mathbf{y}_1[1] & \mathbf{y}_2[1] & \cdots & \mathbf{y}_M[1] \\ & \vdots & & \\ \mathbf{y}_1[k] & \mathbf{y}_2[k] & \cdots & \mathbf{y}_M[k] \\ & \vdots & & \\ \mathbf{y}_1[N-1] & \mathbf{y}_2[N-1] & \cdots & \mathbf{y}_M[N-1] \end{bmatrix}, \quad (1)$$

where $\mathbf{y}_i$ is the i-th signal. Another assumption is that the NBI of a certain signal band is corrupted with zero-mean Gaussian noise, which is uncorrelated with the other frequencies. The equation below describes the corruption of the k-th frequency bin.

$$\mathbf{y}_i[k] = \mathbf{x}[k] + \mathbf{w}_i[k], \quad (2)$$

where $\mathbf{y}_i[k]$ and $\mathbf{x}[k]$ are the discrete Fourier transform (DFT) of the i-th received signal set $\mathbf{y}_i$ and original time synchronization signal $\mathbf{x}$, respectively.

In addition, $\mathbf{w}_i[k]$ denotes the DFT of the i-th zero-mean Gaussian noise. Since we have multiple signal sets, we can estimate the mean value of $\mathbf{y}_i$ to estimate the noise variance in each frequency. The mean estimation can be described as

$$\hat{\mathbf{m}}_y[k] = \frac{1}{M}\sum_{i=1}^{M}\mathbf{y}_i[k], \quad (3)$$

where $\hat{\mathbf{m}}_y$ is the mean estimation of $\mathbf{y}_i$, and M is number of signal sets. With this mean estimation, the covariance matrix can be built as

$$\mathbf{\Sigma} = E_i[(\mathbf{y}_i - \hat{\mathbf{m}}_y)(\mathbf{y}_i - \hat{\mathbf{m}}_y)^T], \quad (4)$$

where $\mathbf{\Sigma}$ is the covariance matrix of noise, and $E_i$ is the expectation with respect to signal set index i. Since the assumed Gaussian noise between frequencies is uncorrelated, the element of covariance matrix $\mathbf{\Sigma}$ would be

$$\begin{cases} \Sigma_{lm} = 0 & \text{if } l \neq m \\ \Sigma_{lm} = \sigma_k^2 & \text{if } l = m = k \end{cases}, \quad (5)$$

where $\Sigma_{lm}$ is the l-th row and m-th column element of matrix $\mathbf{\Sigma}$. Using the noise variance information of each frequency, a frequency with a low SIR can be suppressed to obtain an accurate time delay with higher detection probability.

## 3. FREQUENCY DOMAIN BASED MAXIMUM LIKELIHOOD TIME DELAY ESTIMATOR

Since we obtain the estimated covariance, the maximum likelihood estimator can be applied to obtain the time shift of the signal. If we shift the original time domain signal by $\tau$ as $\mathbf{x}(n-\tau)$, the original time synchronization signal in the frequency domain will appear as

$$\sum_{n=0}^{N-1}\mathbf{x}(n-\tau)e^{-j\frac{2\pi kn}{N}} = \mathbf{x}[k]e^{j\frac{2\pi k\tau}{N}} = \mathbf{x}[k,\tau]. \quad (6)$$

Since we estimate the covariance matrix of $\mathbf{w}_i[k]$, the probability function from the mean and variance from (3) and (4) can be described as

$$p_M(\mathbf{y}_i \mid \mathbf{x},\tau,\sigma_k) = \frac{1}{\sqrt{2\pi}\sigma_k}\exp\left(-\frac{(\mathbf{y}_i[k]-\mathbf{x}[k,\tau])(\mathbf{y}_i[k]^*-\mathbf{x}[k,\tau]^*)}{2\sigma_k^2}\right) \quad (7)$$

Thus, taking the log, expunging unnecessary terms, and summing up all frequencies, we have the following maximum likelihood estimator,

$$L_M(\tau \mid \mathbf{Y},\mathbf{x},\mathbf{\Sigma}) = \sum_{i=1}^{M}\sum_{k=1}^{N}\ln\left(p_M(\mathbf{y}_i \mid \mathbf{x},\tau,\sigma_k)\right)$$
$$= -\sum_{i=1}^{M}\sum_{k=1}^{N}\frac{1}{2\sigma_k^2}(\mathbf{y}_i[k]-\mathbf{x}[k,\tau])(\mathbf{y}_i[k]^*-\mathbf{x}[k,\tau]^*) \quad (8)$$

where * indicates a conjugate. The most probable time delay $\tau$ can thus be described as

$$\hat{\tau} = \arg\max_{\tau}\left(L_M(\tau \mid \mathbf{Y},\mathbf{x},\mathbf{\Sigma})\right), \quad (9)$$

where $\hat{\tau}$ is the estimated time delay.

## 4. TIME DELAY ESTIMAION WITH VON-MISES DISTRIBUTION

Since the variance in the likelihood value of FDMLTDE in (19) does not decrease as $\sigma_k^2$ increases, the detection probability deteriorates in a low SIR environment. This problem is mentioned in section 5. To overcome this problem, we need a scheme that decreases the variance of the likelihood value at a low SIR frequency. Therefore, a von-Mises distribution [6] was adapted to project all values into a unit circle and express only phase data. To analyze the phase data, we decomposed the received frequency domain data into the angle and magnitude. Since the DFT value of $\mathbf{y}_i$ is complex, it can be described as

$$\mathbf{y}_i[k] = r_{i,k}\left(\cos(\theta_{i,k}) + j \cdot \sin(\theta_{i,k})\right)$$
$$\mathbf{x}[k,\tau] = r_0\left(\cos\left(\theta_{0,k} + j\frac{2\pi k\tau}{N}\right) + j \cdot \sin\left(\theta_{0,k} + j\frac{2\pi k\tau}{N}\right)\right). \quad (10)$$

To keep the notation uncluttered, let $\theta_{0,k}^\tau$ be

$$\theta_{0,k}^\tau = \theta_{0,k} + j\frac{2\pi k\tau}{N} \quad (11)$$

Since a von-Mises distribution is only dependent on the angle, we set $r_{i,k} = 1$ to project all values into a unit circle, which results in the normalization of $\mathbf{y}_i[k,\tau]$. If we substitute (10) into equation (7), we now have the equation below, which is described using $r_{i,k}, r_{0,k}, \theta_{i,k}$, and $\theta_{0,k}$.

$$p_M(\theta_{i,k} \mid \theta_{0,k}^\tau, r_{0,k}) = \frac{1}{2\pi\sigma_k^2}\exp\left(\frac{r_{0,k}}{\sigma_k^2}\cos(\theta_{i,k} - \theta_{0,k}^\tau) - \frac{r_{0,k}^2 + 1}{2\sigma_k^2}\right) \quad (12)$$

If we define $m_k = r_0 / \sigma_k^2$, we then have the von-Mises distribution shown below after removing terms irrelevant to $\theta_{i,k}$.

$$p_M(\theta_{i,k} \mid \theta_{0,k}^\tau, m_k) = \frac{1}{2\pi I_0(m_k)}\exp\left(m_k \cos(\theta_{i,k} - \theta_{0,k}^\tau)\right), \quad (13)$$

where $I_0(m_k)$ is the zeroth-order Bessel function of the first kind. If we build a log likelihood function for the probability function in (13) for every k and all signal set i, it will appear like

$$\Lambda_M(\theta_0^\tau \mid \boldsymbol{\theta}, \mathbf{m}) = \sum_{k=1}^{N}\sum_{i=1}^{M}\ln(p_M(\theta_{i,k} \mid \theta_{0,k}^\tau, m_k)). \quad (14)$$

Since $\theta_{i,k}$ is dependent on $\tau$, the most probable $\hat{\tau}$ can be described as below. In addition, terms irrelevant to $\theta_{i,k}$ were removed.

$$\hat{\tau} = \arg\max_\tau \Lambda_M(\tau \mid \boldsymbol{\theta}, \mathbf{m})$$
$$= \arg\max_\tau \sum_{i=1}^{M}\sum_{k=1}^{N} m_k \cos\left(\theta_{i,k} - \theta_{0,k} - \frac{2\pi k\tau}{N}\right) \quad (15)$$

## 5. VARIANCE OF LIKELIHOOD VALUE PROBLEM

In this section, we describe the likelihood variance problem of FDMLTDE at low SIR frequencies. According to equation (8), the expected difference in the log likelihood value between the true time delay $\tau$ that we search for and the incorrect time delay $\tilde{\tau}$ can then be described as

$$E_i\left[\Delta L_M\right] = L_M(\tau) - L_M(\tilde{\tau}) = \left(\sum_{k \in H}\frac{\sigma_k^2}{2\sigma_k^2} - \sum_{k \in H}\frac{2\sigma_x^2 + \sigma_k^2}{2\sigma_k^2}\right) \quad (16)$$

where the summation over index i is dropped, and $\sigma_x^2$ is the variance of the original time synchronization signal. For a convenient analysis, we consider $\mathbf{x}[k,\tau]$ to be uncorrelated with respect to $\tau$. Since $\sigma_x^2$ is fixed, $E[\Delta L_M]$ will decrease as $\sigma_k^2$ increases. Therefore, if the variance of the likelihood value does not decrease as $\sigma_k^2$ increases, the detection probability will decrease because the likelihood value of false time delay $\tilde{\tau}$ frequently exceeds the likelihood value of the true $\tau$. The variance of the likelihood value of FDMLTDE can be described through the following equation.

$$Var_i\left[L_M(\tau)\right] = \sum_k \frac{2\sigma_k^4}{4\sigma_k^4}$$
$$Var_i\left[L_M(\tilde{\tau})\right] = \sum_k \frac{8\sigma_x^4 + 8\sigma_x^2\sigma_k^2 + 2\sigma_k^4}{4\sigma_k^4} \quad (17)$$

As $\sigma_k^2$ increases, the variance of the likelihood value of false time delay $\tilde{\tau}$ converges, as described in the left side of Fig. 4.

However, the variance of the likelihood value of the proposed TDE in (15) is at maximum, as shown below.

$$Var_i\left[\Lambda_M(\tau)\right]$$
$$= Var_i\left[\sum_k m_k \cos\left(\theta_{i,k} - \theta_{0,k} - \frac{2\pi k\tau}{N}\right)\right] \leq \frac{m_k}{2} = \frac{r_0}{2\sigma_k^2} \quad (18)$$

Since the maximum variance of the likelihood value of the proposed TDE decreases as $\sigma_k^2$ increases, the likelihood value of false $\tilde{\tau}$ exceeds the likelihood value of true $\tau$ less frequently. To show this property through a quantitative measurement, we compared the ratio between the variance of the likelihood value and the expectation of a likelihood gap, as described in right side of Fig. 4. From this view point, we can expect that the proposed TDE will show a better performance especially in a low SIR environment.

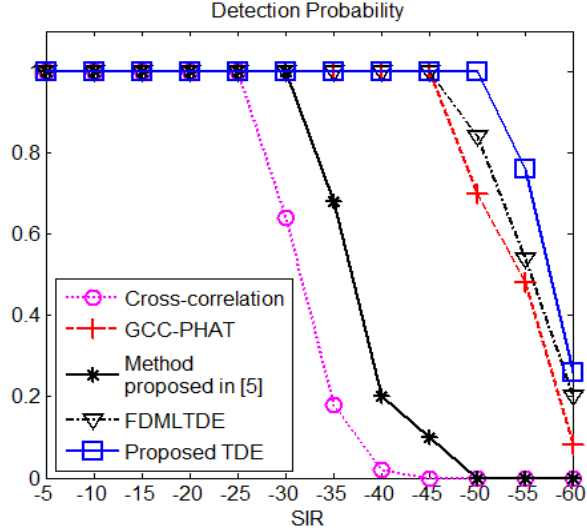

**Fig. 3.** Plot of detection probability for each estimator in a different SIR. (Signal repetition = 5)

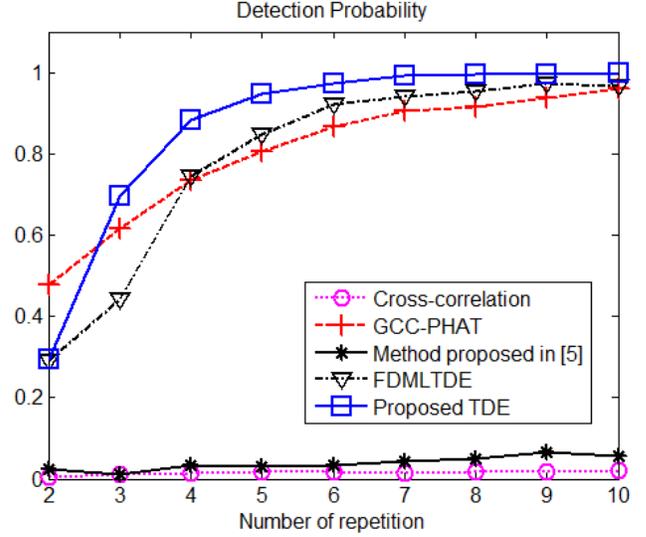

**Fig. 5.** Detection probability plot for each estimator at a different signal repetition (SIR = -50dB).

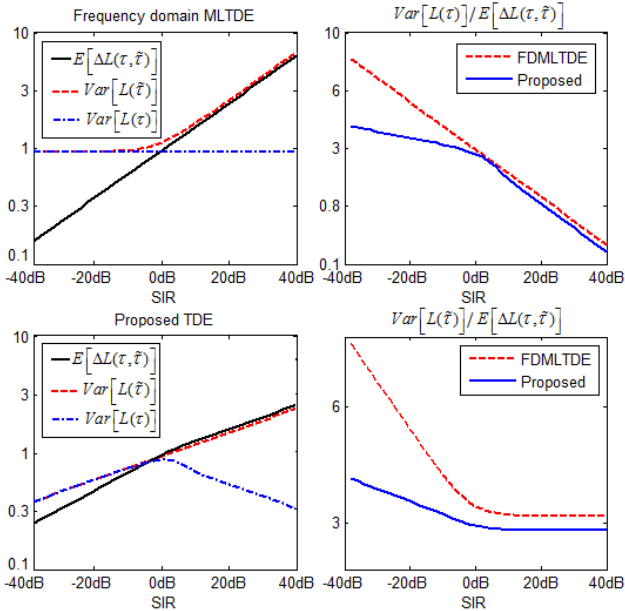

**Fig. 4.** Expectation value for variance of the likelihood value of two estimators.

## 6. NUMERICAL EXPERIMENT RESULT

We tried to verify the performance of our proposed TDE using a Monte-Carlo simulation. We evaluated the performance of the time delay estimator based on the detection probability by considering it as a success if time delay $\tau$ with the maximum likelihood value is equal to the predetermined time delay. Through a communication channel, wide band noise (WBN) equally distributed at every frequency was added with SNR = 0 dB. In addition, NBI noise focusing on a certain frequency band was added.

The normalized interference bandwidth (NBW) can be defined as

$$NBW = BW_I / BW_{Total} = 0.2. \quad (19)$$

The unit synchronization signal length was 512 samples, and a CAZAC sequence [7] was used for the synchronization signal. Fig. 3 shows the detection probability for estimators with a different SIR. We compared the performance of our proposed TDE with a typical cross-correlation method, a FDMLTDE, the method proposed in [5], and a widely used frequency domain TDE, called a generalized cross correlation with phase transform (GCC-PHAT) [8]. Fig. 5 shows the performance of the proposed TDE with different repetition numbers. More than two repetitions have a meaningful suppressing effect on the low SIR band and show a higher detection probability.

## 7. CONCLUSION

In this paper, a methodology used to estimate a time delay with NBI was proposed. By exploiting multiple identical signals, we extracted the noise variance from the signal ensemble. In addition, by employing a von-Mises distribution to reduce the variance of the likelihood value, we reduced the probability of a misdetection of the time delay. Thus, these frequency domain based ML time delay estimators and our proposed time delay method showed a more accurate detection probability in an NBI environment compared to other typical methods. Our proposed technique can be also applied to a positioning system such as GPS or INS, where a robust TDE for NBI is needed.

# 7. REFERENCES

[1] van de Beek, J.-J.; Sandell, M.; Borjesson, P.O., "ML estimation of time and frequency offset in OFDM systems," *Signal Processing, IEEE Transactions on* , vol.45, no.7, pp. 1800-1805, Jul 1997.

[2] Schmidl, T.M.; Cox, D.C., "Robust frequency and timing synchronization for OFDM," *Communications, IEEE Transactions on* , vol.45, no.12, pp.1613,1621, Dec 1997.

[3] Bensley, S.E.; Aazhang, B., "Maximum-likelihood synchronization of a single user for code-division multiple-access communication systems," *Communications, IEEE Transactions on* , vol.46, no.3, pp.392,399, Mar 1998.

[4] Sahmoudi, M.; Amin, M.G., "Improved Maximum-Likelihood Time Delay Estimation for GPS Positioning in Multipath, Interference and Low SNR Environments," *Position, Location, And Navigation Symposium, 2006 IEEE/ION* , vol., no., pp.876,882, April 25-27, 2006

[5] Yuan Tian, Xia Lei, Yue Xiao, and Shaoqian Li, "Time Synchronization for OFDM Systems with Narrowband Interference", *Rough Sets and Knowledge Technology: Lecture Notes in Computer Science* Volume 5589, pp 491-496, 2009.

[6] Christopher M. Bishop, *Pattern Recognition and Machine Learning*, Springer, USA, 2006, pp. 107-109.

[7] Chu, D.: Polyphase codes with good periodic correlation properties. *IEEE Transactions on Information Theory 18(4)*, pp. 531–532 ,1972.

[8] Knapp, C.; Carter, G. Clifford, "The generalized correlation method for estimation of time delay," *Acoustics, Speech and Signal Processing, IEEE Transactions on* , vol.24, no.4, pp.320,327, Aug 1976.